\documentclass[twocolumn,prb,amsmath,amssymb,showpacs,superscriptaddress]{revtex4}

\usepackage{graphicx}

 \DeclareMathOperator{\Tr}{\mathrm{Tr}}
 \DeclareMathOperator{\Real}{\mathrm{Re}}

\begin{document}

 \title{Minigap in superconductor--ferromagnet junctions with inhomogeneous magnetization}

 \author{D.~A.~Ivanov}
 \affiliation{Institute of Theoretical Physics, \'Ecole Polytechnique F\'ed\'erale de Lausanne (EPFL), CH-1015 Lausanne,
              Switzerland}

 \author{Ya.~V.~Fominov}
 \affiliation{L.~D.~Landau Institute for Theoretical Physics RAS, 119334 Moscow, Russia}

\date{18 June 2006}

\begin{abstract}
We consider the minigap in a disordered ferromagnet (F) in contact with a superconductor (S) in the situation when the
magnetization of the F layer is inhomogeneous in space and noncollinear. If the magnetization is strongly inhomogeneous,
it effectively averages out, and the minigap survives up to the exchange field $h_c \sim (L/a) E_\mathrm{Th}$, where $L$
is the thickness of the F layer, $a$ is the scale on which the magnetization varies, and $E_\mathrm{Th}$ is the Thouless
energy. Technically, we use the ``triplet'' version of the Usadel equations, including both singlet and triplet
components of the Green's functions. In many cases, the effect of disordered magnetization may be effectively included
in the conventional Usadel equations as the spin-flip scattering term. In the case of low-dimensional magnetic
inhomogeneities (we consider spiral magnetization as an example), however, the full set of ``triplet'' equations must be
solved.
\end{abstract}






\maketitle

\section{Introduction}

A normal metal in contact with a superconductor acquires some superconductive properties; this phenomenon is known as
the proximity effect. One of the most prominent manifestations of the proximity effect is the (mini)gap in the
single-electron spectrum of the normal metal. While the density of states in the normal metal is spatially dependent,
the minigap is a property of the normal part as a whole. In the diffusive limit, if the thickness of the normal metal
$L$ is larger than the coherence length, the characteristic scale of the minigap $E_g$ is set by the Thouless energy
$E_\mathrm{Th} = D / L^2$, where $D$ is the diffusion constant (we put $\hbar=1$).\cite{GK} Experimentally, the minigap
can be directly probed by a scanning tunneling microscope (see, e.g., Ref.~\onlinecite{Moussy} and references therein).

Nowadays, the proximity effect in more complicated superconductor--ferromagnet (SF) structures is actively studied (see
Ref.~\onlinecite{Buzdin} for a recent review of theoretical and experimental progress). While the majority of earlier
theoretical studies treated the case of single-domain ferromagnets, one of the open questions is the influence of a
domain structure. For example, in some setups, inhomogeneous magnetization may generate long-range triplet
superconducting correlations.\cite{BVE_review}

In a single-domain ferromagnet, the exchange field $h$ (measured in energy units) shifts the densities of states for the
two spin subbands in the opposite directions, therefore the minigap in the spectrum closes at $h \sim
E_\mathrm{Th}$.\cite{FL} A domain structure would effectively lead to averaging the nonuniform field $\mathbf h(\mathbf
r)$ acting on electrons; hence one can expect that the minigap would survive even at $h\gg E_\mathrm{Th}$. In previous
studies, the influence of inhomogeneous magnetization on the density of states was considered only in situations where
the minigap was absent,\cite{BVEmanif,VFE} while the influence of a domain structure on the minigap has not been
investigated.

In this paper, we study the minigap in SF junctions in the presence of inhomogeneous magnetization with the help of the
Usadel equations in the form allowing for the triplet superconducting component.\cite{BVE_review} We introduce a
generalization of the well-known $\theta$ parametrization,\cite{Belzig} which involves, in addition to $\theta$, a
complex vector function $\mathbf M$ with the same number of components as for the exchange field $\mathbf h$ (three in
the general case).

The problem has three relevant energy scales: $E_\mathrm{Th}$, $h$, and the energy $E_d=D/a^2$ associated with the
length scale $a$ over which the magnetization varies (typically, of the order of the domain size). Note that the
superconducting gap $\Delta$ will not play any role, as it is taken to be much larger than $E_\mathrm{Th}$ and $h$.
Throughout the paper we assume that the domains are small and that the ferromagnetic exchange field is weak,
\begin{equation} \label{cond1}
E_d \gg h, \qquad E_d \gg E_\mathrm{Th},
\end{equation}
making no assumption on the relative scale of $h$ and $E_\mathrm{Th}$.

\begin{figure}
 \includegraphics[width=0.95\hsize]{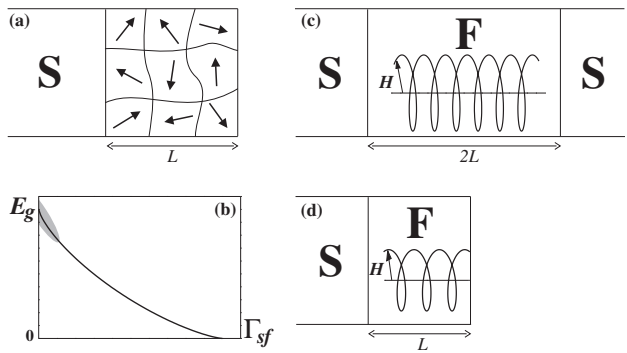}
\caption{(a)~SF junction with disordered magnetization. (b)~Schematic dependence of the minigap $E_g$ on the effective
spin-flip rate $\Gamma_{sf}$ (see Ref.~\onlinecite{CBI} for details); the spin-flip approximation is not applicable in
the shaded region (see the text for details). (c) and (d)~SFS and SF junctions with spiral magnetic order.}
 \label{fig}
\end{figure}

First we consider a SF (or SFS) system with randomly oriented magnetization [Fig.~\ref{fig}(a)] described by the pair
correlation function
\begin{equation} \label{h-corr}
\overline {\mathbf h (\mathbf r) \mathbf h (\mathbf r')} = F(\mathbf r - \mathbf r'),
\end{equation}
where the averaging is over an ensemble. We assume that the integral of the correlation function $F(\mathbf r - \mathbf
r')$ vanishes,
\begin{equation} \label{zero-averaging}
\int F({\mathbf r})\, d{\mathbf r}=0,
\end{equation}
and take $a$ to be the corresponding length scale. Physically, it means that the field
$\mathbf h$ averages out on the scale $a$ (of the order of several domain sizes). This assumption appears
necessary for the smallness of the triplet superconducting correlations and for the possibility to take them into
account as an additional effective local term in the Usadel equation on the singlet component. Such a reduction
is possible for the three-dimensional disorder
(in the whole range of parameters, except at very low exchange fields and at energies near the minigap edge, see
discussion below): in this case, the inhomogeneous field $\mathbf h$ may be effectively included
in the conventional Usadel equation for the singlet component of the pairing correlations
as the spin-flip term\cite{AG} with the spin-flip rate
\begin{equation} \label{Gamma_sf}
\Gamma_{sf} = - \frac{\left< \mathbf h \nabla^{-2} \mathbf h \right>}D \sim \frac{h^2}{E_d}
\end{equation}
(the averaging is over a region much larger than $a$). Note that our result is different from that for pointlike
impurities or disorder,\cite{AG,Houzet} because we consider $\mathbf h$ varying on length scales much larger than the
elastic mean free path.

The SF system with constant $\Gamma_{sf}$ in the F part has already been studied in detail.\cite{CBI} The critical value
of $\Gamma_{sf}$ where the gap closes is  $\Gamma_{sf} / E_\mathrm{Th}=\pi^2/16$, which translates into the critical
value of the exchange field
\begin{equation} \label{h-crit}
h_c \sim \sqrt{E_\mathrm{Th} E_d} \sim  \frac{L}{a} \,  E_\mathrm{Th} .
\end{equation}

If the disorder is one- or two-dimensional, one cannot easily reduce the nonhomogeneous field $\mathbf h$ to a spin-flip
term and needs to solve the full nonlinear set of equations. As an illustration we consider magnetic systems with the
spiral magnetization [Figs.~\ref{fig}(c) and \ref{fig}(d)]. Remarkably, the critical values of the exchange fields for
those systems also have the order of magnitude (\ref{h-crit}).

\section{Method}

The Usadel equation for the Green's function $\check g$ (which is a $4\times 4$ matrix in the Nambu and spin spaces
satisfying the normalization condition $\check g^2 = \check 1 \equiv \hat\tau_0 \hat\sigma_0$) has the
form\cite{BVE_review,comment}
\begin{equation} \label{Usadel}
D \nabla \left( \check g \nabla \check g \right) + i E \left[ \hat\tau_3 \hat\sigma_0, \check g \right] - i \left[
\hat\tau_3 (\mathbf h \hat{\boldsymbol\sigma}) , \check g \right] - \Delta \left[ \hat\tau_1 \hat\sigma_0 , \check g
\right] =0,
\end{equation}
where $\hat\tau$ and $\hat\sigma$ are the Pauli matrices in the Nambu and spin spaces, respectively; $E$ is the energy;
$\mathbf h$ is the exchange field; and $\Delta$ is the superconducting order parameter. For simplicity, we choose
$\Delta$ to be real and do not consider systems with a phase difference.

The solution has the form
\begin{equation} \label{g_components}
\check g = \hat\tau_3 \left( g_0 \hat\sigma_0 + \mathbf g \hat{\boldsymbol\sigma} \right) + \hat\tau_1 \left( f_0
\hat\sigma_0 + \mathbf f \hat{\boldsymbol\sigma} \right).
\end{equation}
The normalization condition can be resolved by the parametrization
\begin{align}
g_0 &= M_0 \cos\theta, & \mathbf g &= i \mathbf M \sin\theta, \notag \\
f_0 &= M_0 \sin\theta, & \mathbf f &= -i \mathbf M \cos\theta, \label{parametrization}
\end{align}
with complex functions $\theta$, $M_0$, and $\mathbf M$, and with the constraint
\begin{equation}
M_0^2 - \mathbf M^2 = 1.
\end{equation}
The Usadel equation (\ref{Usadel}) then yields one scalar and one vector equation:
\begin{align}
& \frac D2 \nabla^2 \theta + M_0 \left( i E \sin\theta + \Delta \cos\theta \right) - ( \mathbf{h M} ) \cos\theta =0,
\label{Us1} \\[10pt]
& \frac D2 \left( \mathbf M \nabla^2 M_0 - M_0 \nabla^2 \mathbf M \right) - \mathbf M ( i E \cos\theta - \Delta
\sin\theta )
\notag \\
& - \mathbf h M_0 \sin\theta =0. \label{Us2}
\end{align}

We may note the following general properties of Eqs.\ (\ref{Us1}) and (\ref{Us2}):

(i) In the absence of ferromagnetism, $\mathbf h =0$, $M_0 =1$, $\mathbf M =0$, and Eqs.\ (\ref{Us1}) and (\ref{Us2})
reduce to the conventional Usadel equation.

(ii) For the uniform nonzero magnetization $\mathbf h ={\rm const}$, Eqs.\ (\ref{Us1}) and (\ref{Us2}) imply the triplet
vector $\mathbf M$ directed along the field $\mathbf h$.\cite{long-range-triplet}

(iii) A convenient feature of the parametrization (\ref{parametrization}) is that in the Matsubara representation [with
$-iE$ replaced by the Matsubara frequency $\omega_n$ in Eqs.\ (\ref{Us1}) and (\ref{Us2})] the
functions $\theta$, $M_0$, and $\mathbf M$ are real.

The density of states (summed over spin projections and normalized to the normal-metallic value) is standardly expressed
via the retarded and advanced Green's functions, $\nu = \Tr \left[ \hat\tau_3 \hat\sigma_0 \left( \check g^R - \check
g^A \right) \right] /8$, which yields
\begin{equation}
\nu = \Real g_0 = \Real ( M_0 \cos\theta ).
\end{equation}

Below we consider SF and SFS systems and employ the rigid boundary conditions, which imply that the bulk solution in the
superconductor with constant $\Delta$ is valid up to the SF interface, and the Green's function is continuous. The rigid
boundary conditions are applicable if the SF interface is transparent and the F material is much more disordered than
the S one. Thus at SF interfaces
\begin{equation} \label{bc1}
\theta= \frac\pi 2,\qquad M_0=1,\qquad \mathbf M =0,
\end{equation}
where the first condition is justified since $h\ll \Delta$ and we consider energies $E \lesssim E_{\mathrm{Th}}$, hence
$E \ll \Delta$. The boundary conditions at the free surface of the F layer in the SF system are
\begin{equation} \label{bc2}
\frac{d\theta}{dz} = 0,\qquad \frac{d M_0}{dz} =0,\qquad \frac{d \mathbf M}{dz} =0.
\end{equation}
Below we consider Eqs.\ (\ref{Us1}) and (\ref{Us2}) in the F part where $\Delta=0$; the superconducting
correlations are induced due to the boundary conditions (\ref{bc1}).

\section{Disordered magnetization}

We start our analysis with the disordered system [Fig.~\ref{fig}(a)]. To reduce the effect of the inhomogeneous field
$\mathbf h$ to a spin-flip term, we use the self-consistent scheme for determining the spin-flip rate $\Gamma_{sf}$: We
first solve the conventional ``spin-flip'' Usadel equation,
\begin{equation} \label{Us_sf}
\frac D2 \nabla^2 \theta + i E \sin\theta - 2 \Gamma_{sf} \sin\theta \cos\theta =0.
\end{equation}
We expect that variations of $\mathbf h$ in space lead to the effective averaging of the magnetization, hence the vector
part $\mathbf M$ of the Green's function is small. Thus we substitute the solution of Eq.\ (\ref{Us_sf}) into the
linearized version of the vector Usadel equation (\ref{Us2}) to obtain
\begin{equation} \label{2lin}
\left( \frac D2 \nabla^2 + i E \cos\theta \right) \mathbf M = - \mathbf h \sin\theta.
\end{equation}
This equation is solved for $\mathbf M$ with the appropriate boundary conditions, and further $\Gamma_{sf}$ is found
self-consistently as
\begin{equation} \label{Gamma_sf1}
\Gamma_{sf} = \left< \frac{\mathbf{hM}}{2 \sin\theta} \right>,
\end{equation}
where the average is taken over a region much larger than the characteristic scale of variation of the field $\mathbf
h$.

For this scheme to work, not only must we require $|\mathbf M| \ll 1$, but the (typically stronger) condition
\begin{equation}
|\mathbf M |^2 \ll \frac{\Gamma_{sf}}{E_\mathrm{Th}}
\end{equation}
[since in the scalar Usadel equation (\ref{Us1}) we neglect the term quadratic in $\mathbf M$, while keeping
$\Gamma_{sf}$].

Equation (\ref{2lin}) is an inhomogeneous linear equation on $\mathbf M$. We consider energies $E \lesssim
E_\mathrm{Th}$, hence $|\theta| \sim 1$, and the $(i E \cos\theta)$ term affects the Green's function of the linear
operator in the left-hand side at a length scale of order $L$. It can be neglected if it produces a small relative
correction to $\mathbf M$, which in turn depends on the number of dimensions in which $\mathbf h(\mathbf r)$ varies.
While we always consider three-dimensional (3D) samples, $\mathbf h$ can either depend on all the three coordinates or
be a function of only two (quasi-2D films) or one (quasi-1D wires) coordinate. Without the $(i E \cos\theta)$ term, the
Green's function of Eq.\ (\ref{2lin}) is  $G \propto 1/r$ in 3D, $G \propto \ln r$ in 2D, and $G \propto r$ in 1D. The
inaccuracy introduced due to neglecting the $(i E \cos\theta)$ term is
\begin{equation}
\delta \mathbf M(\mathbf r) = -\int \delta G(\mathbf r- \mathbf r_1) \mathbf h(\mathbf r_1) \sin\theta (\mathbf r_1) d^d
\mathbf r_1,
\end{equation}
where the correction $\delta G$ is negligible at length scales smaller than $L$. A straightforward estimate using the
correlation function (\ref{h-corr}) with the vanishing integral (\ref{zero-averaging}) gives
$\overline{|\delta \mathbf M|^2} \sim a^{2+d} L^{2-d} h^2/D^2$ and $\overline{|\mathbf M|^2} \sim a^4 h^2/D^2$, hence
\begin{equation}
\frac{\overline{|\delta \mathbf M|^2}}{\overline{|\mathbf M|^2}} \sim \left( \frac aL \right)^{d-2}.
\end{equation}
Therefore we can neglect the $(i E \cos\theta)$ term  in 3D, can obtain order-of-magnitude estimates in this way in 2D,
and cannot do it in 1D. Note that we have also neglected the effect of boundaries. It can be taken into account via
``mirror charges'' (of the same or opposite sign for the boundaries at which $d\mathbf M /dx =0$ or $\mathbf M =0$,
respectively), and can be neglected under the same conditions.

Now we assume that the disorder is three-dimensional and neglect the $(i E \cos\theta)$ term in Eq.\ (\ref{2lin}).
Further, since $\mathbf h$ varies in space much faster than $\theta$, we find
\begin{equation}
\mathbf M =- \frac 2D \left( \nabla^{-2} \mathbf h \right) \sin\theta,
\end{equation}
and Eq.\ (\ref{Gamma_sf1}) immediately leads to the effective spin-flip scattering rate (\ref{Gamma_sf}). That
expression may be interpreted as the electrostatic energy of the ``charges'' $\mathbf h$. Estimating $\Gamma_{sf} \sim
h^2 a^2 /D$ and $|\mathbf M|^2 \sim h^2 a^4/D^2$ translates our assumptions on the smallness of $\mathbf M$ into the
conditions (\ref{cond1}), which thus confirm our linearization in $\mathbf M$.

However, to finally justify the spin-flip approximation, we must avoid one more danger: the noninvertibility of the
``Hamiltonian'' $-\bigl[ (D/2) \nabla^2 + iE \cos\theta \bigr]$. The calculations above were made without taking this
possibility into account. At the same time, this happens exactly at the gap edge at $\Gamma_{sf}=0$ (i.e., at zero
exchange field $\mathbf{h}$). Indeed, below the gap $\theta = \pi/2 + i\vartheta$, where $\vartheta$ is real.
Parametrizing the solutions of the Usadel equation at $\Gamma_{sf}=0$ by $\vartheta_0$, the value of $\vartheta$ at the
free F surface in the SF junction or in the center of the F layer in the SFS junction, and taking into account that the
normal-metallic minigap\cite{GK} $E_g = 0.78 E_\mathrm{Th}$ is the maximum value of $E(\vartheta_0)$, we see that $d
\vartheta / d\vartheta_0$ is the zero-energy eigenfunction of the above ``Hamiltonian''.

So in the vicinity of the point  $\Gamma_{sf}=0$, $E=E_g$, the spin-flip approximation breaks down. In this region, the
zero mode becomes a low-energy mode with a small nonzero energy $E_0$. Expanding $\mathbf M = \sum c_i \mathbf M_i$ in
the basis of the normalized eigenfunctions $\mathbf M_i$ of the ``Hamiltonian'' on the left-hand side of Eq.\
(\ref{2lin}), we find that the contribution of the low-energy mode is determined by
\begin{equation}
c_0 = \frac 1{E_0} \int d^d \mathbf r (\mathbf{hM}_0) \sin\theta.
\end{equation}
Using the correlation function (\ref{h-corr}), we estimate $\overline{|c_0|^2} \sim a^{2+d} L^{-2} h^2/E_0^2$. Finally,
the condition $\overline{| c_0 \mathbf M_0|^2}  \ll  \Gamma_{sf}/E_\mathrm{Th}$ that the contribution from the
low-energy mode is small, yields
\begin{equation} \label{cond2}
\frac{E_0}{E_\mathrm{Th}} \gg \left( \frac aL \right)^{d/2}.
\end{equation}
The energy $E_0$ grows as we go away from the gap-closing point at $\Gamma_{sf}=0$, $E =E_g$. The perturbation theory
yields
\begin{equation} \label{E_0}
\frac{E_0}{E_\mathrm{Th}} \sim \max\left(\sqrt{\frac{|E-E_g|}{E_\mathrm{Th}}}\, ,\, \frac{\Gamma_{sf}}{E_\mathrm{Th}}
\right)\, .
\end{equation}
The combination of Eqs.\ (\ref{cond2}) and (\ref{E_0}) determines a small region near the $\Gamma_{sf}=0$, $E=E_{g}$
point [shown as the shaded region in Fig.~\ref{fig}(b)], where the spin-flip approximation (with the effective rate
$\Gamma_{sf}$) breaks down due to the noninvertibility of the linear operator in Eq.\ (\ref{2lin}).

Summarizing, the Usadel equations (\ref{Us1}) and (\ref{Us2}) in the 3D problem reduce to the conventional Usadel
equation (\ref{Us_sf}) with the spin-flip rate (\ref{Gamma_sf}) if the conditions (\ref{cond1}) and (\ref{cond2}) are
satisfied. Then the results of Ref.~\onlinecite{CBI} for the $E_g(\Gamma_{sf})$ dependence apply. The critical value of
the exchange field, at which the minigap vanishes, is given by Eq.\ (\ref{h-crit}) and is much larger than $h_c$ in the
case of a homogeneous ferromagnet.

\section{Spiral magnetization}

The above discussion in terms of the spin-flip approximation applies to the 3D case and should give an
order-of-magnitude estimate in the 2D case. In 1D setups (with the field $\mathbf h$ depending only on one of the
coordinates), one generally needs to solve the full nonlinear set of Eqs.\ (\ref{Us1}) and (\ref{Us2}). Below we present
an analysis of the two examples [SFS and SF systems, see Figs.~\ref{fig}(c) and \ref{fig}(d), respectively] with the
\textit{regular} spiral magnetic structure:
\begin{equation}
\mathbf h = h (\cos kz, \sin kz, 0),
\end{equation}
where the $z$ direction is perpendicular to the SF interface(s). In
both the examples we are unable to analytically find the full dependence of the minigap on the strength of the field
$h$, but we calculate the critical value $h_c$ at which the minigap closes.

Technically, the SF case turns out to be more complicated, because the linearization over $\mathbf M$ does not work in
this situation (while it is still applicable in the SFS system).

\subsection{SFS case}

\begin{figure}
 \includegraphics[width=0.95\hsize]{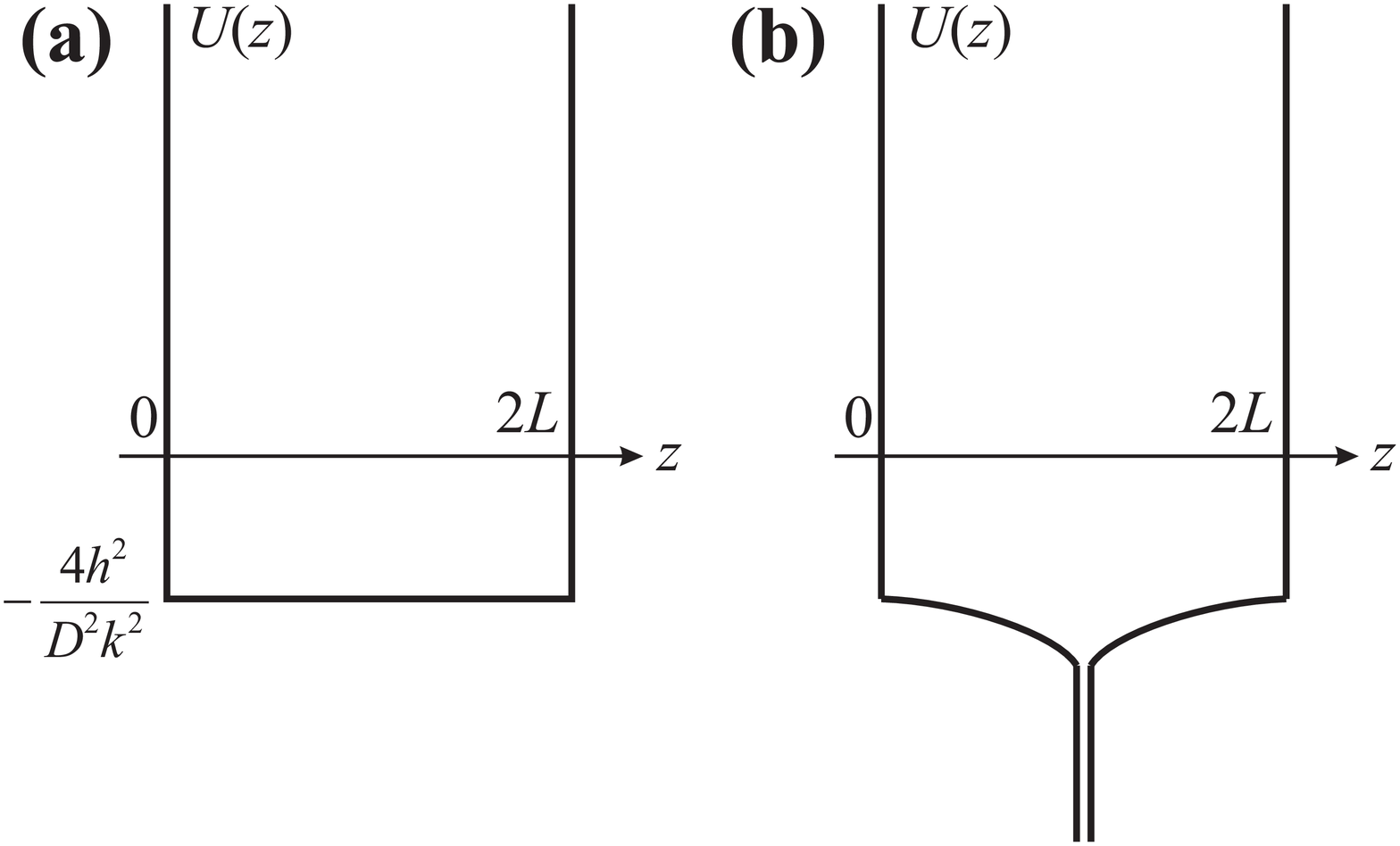}
\caption{Effective quantum-mechanical potentials for (a)~SFS and (b)~SF junctions. The minigap vanishes when the
Hamiltonian with the corresponding potential has the ground state with zero energy. In both cases, the impenetrable
walls are at $z=0$ and $z=2L$. (a)~Rectangular potential well. (b)~Potential well with a $\delta$-functional
contribution in the middle.}
 \label{fig:potentials}
\end{figure}

We consider the spiral ferromagnet of length $2L$ [Fig.~\ref{fig}(c)] with the rigid boundary
conditions (\ref{bc1}) at the SF
interfaces. With these boundary conditions and at zero energy it is possible to perform a derivation similar to that in
the disordered 3D case. To find the gap-closing point, we find the minimal value of $h$ at which the $E=0$ solution with
$\theta=\pi/2$ has a bifurcation (a gapless solution forks out). Linearizing in $(\theta-\pi/2)$ and in $\mathbf M$ at
$E=0$, we solve Eq.\ (\ref{2lin}), without taking into account boundary conditions, by the oscillating function
\begin{equation}
\mathbf M_\mathrm{osc} = \frac{2 \mathbf h}{D k^2}.
\end{equation}
The boundary conditions $\mathbf M (0)= \mathbf M (2L)=0$ are easily satisfied by adding a smooth (nonoscillating) term
linear in $z$ of the same order of magnitude: $\mathbf M =\mathbf M_\mathrm{osc} +\mathbf M_\mathrm{sm}$. On
substituting this solution into the linearized scalar Usadel equation, we find the effective equation for the smooth
part of $\theta(z)$ [the oscillating part of $\theta(z)$ has a smaller order of magnitude and is neglected; only
$\mathbf M_\mathrm{osc}$ needs to be taken into account at this step]:
\begin{equation} \label{Sch}
\left( \nabla_z^2 + \frac{4 h^2}{D^2 k^2} \right) \left( \theta - \frac{\pi}{2} \right) =0
\end{equation}
with the boundary conditions
\begin{equation} \label{bc}
\theta(0)=\theta(2L)= \frac\pi 2.
\end{equation}
Equation (\ref{Sch}) can be viewed as the Schr\"odinger equation at zero energy with $\hbar =1$, mass $1/2$, and the
constant potential $-4h^2/(D^2 k^2)$, while $(\theta-\pi/2)$ plays the role of the eigenfunction. The boundary conditions
(\ref{bc}) correspond to the impenetrable walls at $z=0$ and $z=2L$, see Fig.~\ref{fig:potentials}(a).

Therefore, the bifurcation of a nontrivial solution for the problem (\ref{Sch}) and (\ref{bc})
(i.e., closing of the minigap) with increasing $h$ corresponds to the ground-state energy crossing zero.
This immediately leads to the result
\begin{equation} \label{h_c_SFS}
h_c^\mathrm{SFS}= \left( \frac\pi 4 \right) \frac{D k}L .
\end{equation}
We verify that the linearization over $\mathbf M$ is justified under the conditions (\ref{cond1}) (with $E_d=D k^2$)
which, in turn, reduce to the single requirement $kL\gg 1$ at $h\sim h_c^\mathrm{SFS}$. Note that Eq.\ (\ref{h_c_SFS})
formally coincides with the expression (\ref{Gamma_sf}) in combination with the critical value of $\Gamma_{sf}$ from
Ref.~\onlinecite{CBI}. The critical value of the exchange field is, as in the previous example, much larger
than the Thouless energy: $h_c^\mathrm{SFS} \sim (kL) E_\mathrm{Th}$.

\subsection{SF case}

Another example we consider is one half of the above SFS junction: the SF junction with the ferromagnet of length $L$
[Fig.~\ref{fig}(d)]. In this case, the open boundary conditions cannot be satisfied at small $\mathbf M$ and the
linearization over $\mathbf M$ does not work. [Physically, $\mathbf M$ is enhanced because electrons incident on the
outer F surface through the field $\mathbf h (z)$, feel the same $\mathbf h (z)$ after reflection, hence the effective
average of $\mathbf h$ is nonzero.] Thus we need to solve the full nonlinear Eq.\ (\ref{Us2}) at $E=0$. Below the gap,
$\mathbf M$ is real, and it is convenient to introduce the new \textit{complex} function
\begin{equation}
m(z)= M_1 + i M_2
\end{equation}
(where $M_1$ and $M_2$ are the two components of the vector $\mathbf M$). Then in the new complex notation Eq.\
(\ref{Us2}) takes the form
\begin{equation} \label{m-nonlinear}
\frac D2 \left[ m'' - \left( \frac{M_0''}{M_0} \right) m \right] + h e^{ikz} = 0 , \quad M_0=\sqrt{1+m m^*},
\end{equation}
with the boundary conditions $m(0)=0$ and $m'(L)=0$. We separate rapidly oscillating modes,
\begin{equation} \label{expansion}
m=b_0 + b_1 e^{ikz} + b_{-1} e^{-ikz} + \dots ,
\end{equation}
and keep only the leading ones (the amplitudes $b_n$ are slow functions of $z$ and decrease in magnitude with increasing
$|n|$, as we verify below). The equations for different modes are coupled, and after a straightforward algebra, we
express the amplitudes $b_1$ and $b_{-1}$ via the smooth part $b_0$:
\begin{equation}
b_1= \frac h{D k^2} \left( 2+ |b_0|^2 \right) , \qquad b_{-1}= \frac h{D k^2} b_0^2 .
\end{equation}
Substituting these amplitudes in Eq.\ (\ref{m-nonlinear}) and parametrizing $b_0= -i e^{ikL} \sinh \varphi$, we arrive
at the effective equation on $\varphi(z)$:
\begin{equation}
\varphi'' + \frac{\varkappa^2}2 \sinh\varphi \cosh\varphi=0,
\end{equation}
with the boundary conditions
\begin{equation}
\varphi(0)=0 , \qquad \varphi'(L)=\varkappa \cosh\varphi(L) ,
\end{equation}
where $\varkappa=2h/(Dk)$. The solution to this equation must be substituted back into the scalar Usadel equation
(\ref{Us1}) to find the bifurcation point of the solution $\theta=\pi/2$. Linearizing the equation in the vicinity of
the bifurcation point, we obtain
\begin{multline}
\biggl[\nabla_z^2 + \varkappa^2 \left( 1+\frac{\sinh^2 \varphi}2 \right) -
\left( k\varkappa \sinh \varphi \right) \sin k(z-L)
\\
- \left(\varkappa^2\; \frac{\sinh^2 \varphi}{2} \right) \cos 2k(z-L) + \dots \biggr] \left( \theta-\frac\pi 2 \right) =0.
\end{multline}
Again, we need to take into account not only the smooth part of $\theta$, but also the oscillating modes at $\pm k$:
\begin{equation} \label{expansion1}
\theta - \frac\pi 2 = t_0 + t_{1} \sin k(z-L)  + \dots
\end{equation}
Solving coupled equations for the amplitudes $t_n$ (which are slow functions of $z$ and decrease in magnitude with
increasing the order of the harmonic), we find, to the leading order in $\varkappa/k$,
\begin{equation}
t_{1} = -\left(\frac\varkappa k \sinh \varphi \right) t_0,
\end{equation}
and finally obtain the equation for the smooth component $t_0$:
\begin{equation} \label{Sch1}
\left( \nabla_z^2 + \varkappa^2 \cosh^2\varphi \right) t_0 =0,
\end{equation}
with the boundary conditions
\begin{equation} \label{bcc}
t_0(0)= 0, \qquad t_0'(L) = \Bigl[ \varkappa \sinh \varphi(L) \Bigr] t_0(L) .
\end{equation}
At the bifurcation point (i.e., when the minigap closes), this homogeneous linear equation acquires a nonzero solution.
The resulting problem again has a quantum-mechanical analogy if we symmetrically continue the potential $-\varkappa^2
\cosh^2\varphi (z)$ in the Schr\"odinger equation (\ref{Sch1}) from the $(0,L)$ interval to $(L,2L)$; see
Fig.~\ref{fig:potentials}(b). As a result, we obtain the Schr\"odinger equation at zero energy with $\hbar =1$ and mass
$1/2$, while $t_0$ plays the role of the eigenfunction. At $z=0$ and $z=2L$, we obtain the impenetrable walls, while the
boundary condition (\ref{bcc}) at $z=L$ corresponds to the following $\delta$-functional contribution to the potential:
$-2\left[ \varkappa \sinh \varphi(L)\right] \delta(x-L)$.

Therefore, the appearance of a nontrivial solution for the problem (\ref{Sch1}) and (\ref{bcc}) at the bifurcation point
with increasing $h$ corresponds to the situation, in which the ground state of the Schr\"odinger equation has zero
energy. Numerically, we find that this happens at $\varkappa = 0.5955/L$, which translates into the critical
value of the exchange field
\begin{equation}
h_c^\mathrm{SF}= 0.2977 \frac{Dk}L .
\end{equation}
Having found $h_c^\mathrm{SF}$, we can check that $b_0 \sim 1$, $b_{\pm 1} \sim 1/kL$, $b_{\pm 2} \sim 1/ (kL)^2$,
$t_{1} \sim t_0/kL$, etc., confirming the validity of the expansions (\ref{expansion}) and (\ref{expansion1}). Note
that the critical strength of $\mathbf h$ has the same order of magnitude as in the SFS case [and thus also agrees with
the estimate (\ref{h-crit})], but it has a smaller numerical prefactor.

\section{Conclusions}

In conclusion, we have shown that SF and SFS systems with random three-dimensional domain disorder admit an effective
description in terms of the spin-flip rate $\Gamma_{sf}$. We have also considered SF and SFS junctions with spiral
magnetic order, where the spin-flip approximation does not hold. In all the systems considered, the minigap survives up
to the exchange fields of the order (\ref{h-crit}), i.e., much larger than $E_\mathrm{Th}$. For an experimental
observation of the effects described in the present paper, it is crucial to use weak ferromagnets with a very fine
domain structure, so that the main constraints (\ref{cond1}) are satisfied.

\begin{acknowledgments}
Ya.V.F.\ was supported by the RFBR Grants Nos.\ 04-02-16348 and 04-02-08159, the RF Presidential Grant
No.~MK-3811.2005.2, the Russian Science Support Foundation, the Dynasty foundation, the program ``Quantum Macrophysics''
of the RAS, CRDF, and the Russian Ministry of Education.
\end{acknowledgments}


\begin{thebibliography}{99}
\bibitem{comment}
Our Green's function is related to the one used in Ref.~\onlinecite{BVE_review}, $\check g_\mathrm{BVE}$, via a rotation
$\check g = \check U \check g_\mathrm{BVE} \check U^\dagger$ with $\check U = ( 1+ i\hat\tau_3 \hat\sigma_3)
(1-i\hat\sigma_3)/2$ (plus inessential change of sign of $\mathbf h$). We prefer this definition because then the Usadel
equation possesses explicit symmetry with respect to rotations of the exchange field $\mathbf h$.
\bibitem{long-range-triplet}
Note that, in a uniform ferromagnet, $\mathbf M$ parallel to $\mathbf h$ does not produce a ``long-range triplet
proximity effect'' in the terminology of Ref.~\onlinecite{BVE}. For the long-range effect described in
Ref.~\onlinecite{BVE}, one needs to generate a component of $\mathbf M$ perpendicular to $\mathbf h$. For example,
long-range triplet correlations are absent in Ref.~\onlinecite{Eschrig}: in that particular configuration of the
inhomogeneous field, $\mathbf M$ follows the direction of $\mathbf h$.
\end{thebibliography}
\end{document}